# Effects of Coordinative Arm Swing Movements on the Sense of Agency in Walking Sensation Induced by Kinesthetic Illusion

Eifu NARITA[1], Keigo USHIYAMA[1], Izumi MIZOGUCHI[1] and Hiroyuki KAJIMOTO[1]

[1] Department of Informatics, The University of Electro-Communications, Tokyo, JAPAN

(Email: {narita, ushiyama, mizoguchi, kajimoto}@kaji-lab.jp)

**Abstract** --- Kinesthetic illusion can present a sense of movement without actual physical movement of the body, but it often lacks a sense of agency over the movement. We focused on the sensation of walking induced by the kinesthetic illusion and hypothesized that incorporating coordinated arm swing movements as actual actions could enhance the sense of agency over the kinesthetic illusion. In this study, we implemented a system that switches the vibrations of the thighs and ankles back and forth based on arm swing movements and investigated whether the sense of agency over the walking sensation induced by the kinesthetic illusion changes with or without arm swing movements. The results suggest a tendency for the sense of agency to be enhanced when arm swing movements are combined.

Keywords: alternating vibrations, kinesthetic illusion, sense of agency, tendon vibratory stimulation, walking sensation

## 1 Introduction

In VR experiences, the kinesthetic illusion is useful as a technology that can present a sense of movement without actual physical movement. A kinesthetic illusion is an illusion of limb movement or displacement of position induced by low-frequency vibratory stimulation of tendons [1]. This illusion occurs in the direction of the stretch of the stimulated muscle [2]. By applying this technology, it is possible to compactly present a sense of movement without the need for large devices or direct physical actuation.

However, the kinesthetic illusion remains a passive sensation of being moved by the stimulation, lacking a sense of agency. The sense of agency refers to the feeling that one's movements are controlled by oneself and arises when the predicted sensory feedback from an intended movement closely matches the actual perceived sensory feedback [3].

We focused on a different part of the body that moves coordinately with the stimulated part and hypothesized that combining actual movements with the presented kinesthetic illusion could impart a sense of agency.

We focus on a walking scenario. During walking, arm swing movements occur coordinately with leg movements [4]. By treating active arm swing movements as the motor command for walking, the predicted sensation could be linked with the passive sensation of leg movement induced by the illusion, potentially creating a sense of agency.

In this study, we examine whether combining active arm swing movements with the presentation of walking sensations through a kinesthetic illusion can enhance the sense of agency. In the experiments, we conducted a psychological evaluation comparing between cases where the walking sensation was presented using arm swing movement input and cases where it was passively experienced.

## 2 Related Work

Several methods for presenting walking sensations using the kinesthetic illusion have been proposed [5][6][7]. Leonardis et al. [5] utilized the sensation of the body returning to its original position, which is an aftereffect of the illusion when the vibration stops, to recreate knee joint extension and flexion. Duclos et al. [6] reproduced the sensation of walking by providing vibrations mapped temporally to the stretching of muscles involved in the ankle, knee, and hip joints, based on walking kinematics. According to the study by Tapin et al. [7], a good walking sensation can be achieved by applying vibrations to at least the knee and one other joint (either the hip or the ankle).

To enhance the sense of agency for kinesthetic illusion, systems using brain-computer interfaces (BCIs) based on motor imagery input [8] and systems that take the force exerted when trying to move a fixed body part as input [9] have been proposed. In these methods, the body part intended to move and the body part perceiving the illusion are the same.

## 3 EXPERIMENT

Seven male participants aged 21 to 24 took part in the experiment. They evaluated the strength of the illusion, the subjective realism of the walking sensation, the naturalness of the walking sensation, and the sense of agency in two cases: when vibratory stimuli were provided based on arm swing movements and when vibratory stimuli were provided at a fixed interval without arm movement. The experiment was approved by the ethics committee of the authors' institute (H23035).

### 3.1 Apparatus

An overview of the entire system is shown in Fig.1 . Eight vibrators (Acouve Lab Vp4 series Vp410) were placed on the ventral and dorsal sides of the knees and ankles (two on each side) and fixed with supporters. A signal generation software (Cycling '74 & MI7 Max 8) generated a 70 Hz sine wave, which was routed through an audio interface (Roland, OCTA-CAPTURE) and audio amplifiers (FX-AUDIO- FX202A/ FX-36A PRO), driving the vibrators alternately on the ventral and dorsal sides.

Arm swing movements were detected using an HMD (Head Mounted Display, Meta Quest 3) and its controller. Participants wore the HMD and controller, and vibrations were switched when the direction of the arm swing changed. To prevent false detection, the angle between the HMD and the controller in the sagittal plane was calculated, and detection was set to only occur when a certain angle threshold was exceeded. Although only the right-side controller was used to detect arm swing movements, participants wore controllers on both hands. This approach was chosen to ensure accurate detection of vibration switching timing and to avoid errors caused by slight phase differences between the arms, which naturally move in opposite directions.

### 3.2 Conditions

Vibratory stimuli were applied to the knees and ankles, targeting the flexors on the ventral side and the extensors on the dorsal side (Fig.2 (a)). Ventral vibrations induced kinesthetic illusions of extension, while dorsal vibrations

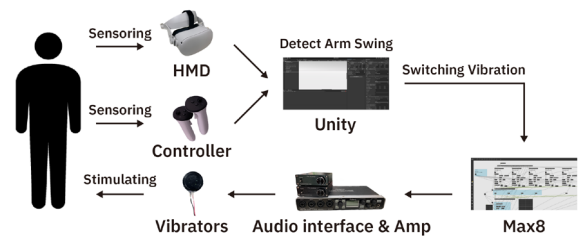

Fig.1 Overview of the system

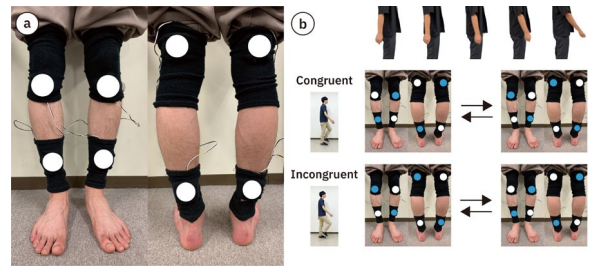

Fig.2 (a) Position of the vibrators, (b) Stimulation conditions with arm swing movements as input induced flexion. Knee extension targeted the quadriceps tendon, and knee flexion targeted the biceps femoris tendon. Similarly, ankle extension targeted the tibialis anterior tendon, and ankle flexion targeted the triceps surae tendon.

The walking sensation was created using alternating vibrations (Fig.2 (b)). Previous research mapped transitions such as ankle extension to flexion and knee extension to flexion [6]. Due to the variations in amplitude and frequency of arm swing movements, a simpler method of alternating states was adopted, with opposite driving sides for the knees and ankles.

Following Naito et al.'s research [10], which reported that 70 Hz induces a strong kinesthetic illusion, this frequency was used in the experiment. It is important to note that dynamically altering the frequency may not enhance the match between arm swing and the illusion of leg movement, as the kinesthetic illusion cannot be fully controlled by adjusting the vibratory frequency alone. The experimental conditions included congruent (right arm swing paired with left leg forward), incongruent (opposite pairing), and passive (constant 1 Hz cycle [6]) conditions. Each set of three conditions was repeated three times, resulting in a total of nine trials. The experimental data utilized the median value of the three repeated measurements, with the order of trials randomized.

### 3.3 Procedure

After positioning the vibrators with supporters, calibration was performed using an accelerometer (Sparkfun, LIS331). The vibration amplitude of each vibrator was adjusted to 80 m/s² at 70 Hz by controlling the volume on the signal generation software.

For each trial, the stimulation duration was set to 60 seconds. In all trials, participants were instructed to swing their arms at their usual walking speed and amplitude. They were also instructed to focus on moving their upper arms. A 30-second rest period was provided between trials.

During stimulation, participants were required to stand and keep their eyes closed. To closely approximate the posture during actual walking, the experiment was conducted in a standing position. Additionally, participants wore headphones playing white noise to block out the sound of the vibrators.

After the stimulation, participants rated the following on a 9-point scale: the strength of the illusion (1: did not feel it at all, 9: felt it very strongly), the realism of the walking sensation (1: did not feel like a walking sensation, 9: felt like real walking), the naturalness of the walking sensation (1: felt very unnatural, 9: felt very natural), and the sense of agency (1: felt like being moved by an external force, 9: felt like moving by oneself).

Comments were collected from participants after the experiment.

## 4 RESULTS

The evaluation results are shown in Fig.3 . The Friedman test was used for each evaluation item, and Bonferroni correction was applied for multiple comparisons. A p-value of 0.05 was used as the significance level for all tests. The statistical testing software used was IBM SPSS Statistics.

No significant differences were observed between the conditions for the strength of the illusion (Fig.3 (a)) and the naturalness of the walking sensation (Fig.3 (c)).

For the realism of the walking sensation (Fig.3 (b)), a significant difference was observed only between the passive condition and the incongruent condition (p = 0.015).

For the sense of agency (Fig.3 (d)), significant differences were observed between the passive condition and the congruent condition (p = 0.048) and between the passive condition and the incongruent condition (p = 0.048).

## 5 DISCUSSION

No significant differences in the strength of the illusion were observed between conditions with arm swing movements and the passive condition (Fig.3 (a)), suggesting that active movements of the arms did not enhance the illusion itself. The realism of the walking sensation was higher in the incongruent condition (Fig.3 (b)), likely influenced by arm swings, with participants noting a dragging feet sensation due to a fixed standing position.

The increased realism in the incongruent condition may be due to a delay in illusion onset or difficulty distinguishing from the congruent condition. The illusion induced by the vibration switch might have been in reverse phase, similar to previous findings on the temporal characteristics of reciprocal kinesthetic illusions induced by vibratory stimuli of different frequencies applied to two antagonistic muscles [11]. However, the lack of significant difference between incongruent and congruent conditions indicates participants likely did not distinguish between them.

No significant differences in naturalness of the walking sensation were found between congruent and incongruent conditions (Fig.3 (c)). Some participants noted differences in naturalness related to coordination. Future studies should investigate individual differences in walking speed and stride length and the effects of arm swings synchronized with periodic leg movements.

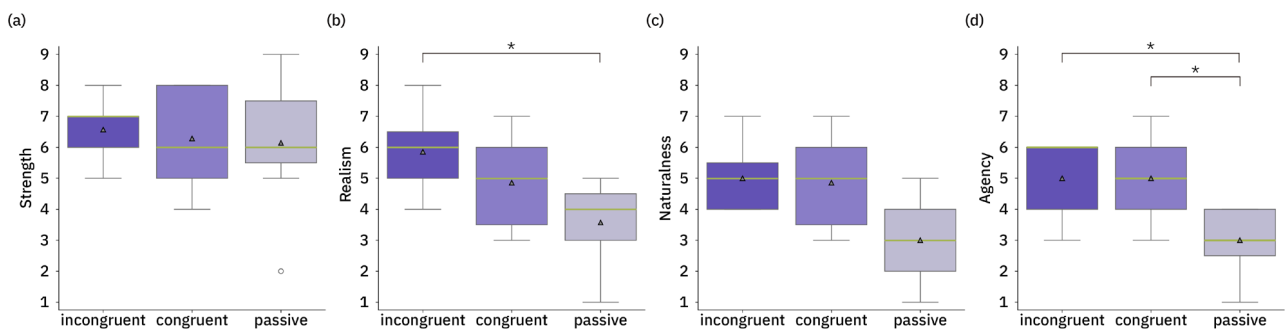

Fig.3 Evaluation of (a) strength of the illusion, (b) realism of the walking sensation, (c) naturalness of the walking sensation, and (d) sense of agency for each condition (incongruent, congruent, passive)

Arm swing movements improved the sense of agency in the walking sensation induced by the kinesthetic illusion (Fig.3 (d)). Participants felt they were moving their legs themselves, possibly due to predicted sensations from arm swings including leg movement. There have been reports of lower limb muscle activity induced by arm swing movements [12], and the functional coupling between the upper and lower limbs might have contributed to the enhanced sense of agency. Further research should explore the effects of other combined actions, such as joystick inputs.

The sense of agency improvement was minimal regardless of whether arm swing movements were in-phase or out-of-phase with leg movements. Previous studies also found no significant difference in agency with first-person avatar movements in VR under similar conditions [13]. The lack of significant differences in naturalness suggests participants may not have distinguished these conditions. Further investigation is needed on the impact of delays in illusion presentation and control of walking speed.

## 6 Conclusion

In this study, we hypothesized that combining coordinated actual movements of body parts different from the stimulated sites could enhance the sense of agency of kinesthetic illusions. We focused on walking accompanied by arm swing movements and conducted psychological evaluations to see if the presence or absence of arm swing movements would affect the sense of agency in the presentation of walking sensations induced by kinesthetic illusions. The results suggest that arm swing movements tend to enhance the sense of agency, regardless of their coordination with leg movements.

In the future, we will investigate the effects of having participants perform arm swing movements in coordination with the presented walking sensations, the impact of combining different actions, and the differences caused by delays in the presentation of the illusion.

## Acknowledgement

This research was supported by JSPS KAKENHI Grant Number JP20H05957.